\documentclass[a4paper]{revtex4}
\usepackage{graphicx}
\usepackage{fancyhdr}
\usepackage{amsmath}
\usepackage{color}

\begin{document}

\title{Extra Higgses at LHC: the EW Road to Baryogenesis}

%

\author{Jose Miguel No}
\affiliation{Department of Physics and Astronomy, University of Sussex, Brighton BN1 9QH, United Kingdom}

\begin{abstract}
A cosmological first order electroweak phase transition could explain the origin of the matter-antimatter asymmetry in the Universe. 
Such a phase transition does not occur in the Standard Model, while it becomes possible with the existence of a second Higgs doublet in Nature.
We obtain the properties of the new scalars $H_0$, $A_0$ and $H^{\pm}$ that lead to such a phase transition, 
showing that its characteristic signature at LHC would be the observation of the decay $A_0 \rightarrow H_0 Z$.
We analyze the LHC search prospects for this decay in the $\ell \ell b\bar{b}$ and $\ell \ell W^{+} W^{-}$ final 
states, showing that either one is promising at the early stages of the 14 TeV run.

\end{abstract}

\maketitle

\thispagestyle{fancy}


\section{Introduction}
A key goal of the Large Hadron Collider (LHC) physics programme is to reveal the properties of the electroweak symmetry breaking (EWSB) sector in Nature. 
The ATLAS and CMS data from the 7 and 8 TeV runs of LHC show that the properties of the newly discovered Higgs particle are compatible with 
those expected for the Standard Model (SM) Higgs boson $h$. Yet, it still needs to be determined whether the scalar sector consists of one $SU(2)_L$ doublet 
or has a richer structure, with additional states. 

A particularly appealing feature of extensions of the SM scalar sector, like Two-Higgs-Doublet-Models (2HDMs), is that they could successfully explain the 
generation of the cosmic matter-antimatter asymmetry via Electroweak Baryogenesis \cite{EWBG}. A key requirement in this context is that the Electroweak Phase 
Transition (EWPT) in the early Universe be strongly first order, which for the SM with a $m_h = 125$ GeV Higgs is not the case \cite{Kajantie:1995kf}.

Here we review the results of \cite{Dorsch:2014qja}, showing that the characteristic signature of a strongly first order 
EWPT in 2HDMs is the decay $A_0 \rightarrow Z H_0$ , and analyzing its promising search prospects at the 
14 TeV run of LHC in the $\ell \ell\,b\bar{b}$ and $\ell \ell\,W^{+} W^{-} \to \ell\ell\ell\ell\nu\nu$ final states.
This signature could then provide a connection between the generation of 
the cosmic matter-antimatter asymmetry in the Early Universe and searches for new physics at LHC.

\section{Two Higgs Doublets and the Electroweak Phase Transition}

The scalar sector of a 2HDM contains two scalar doublets $\Phi_{1,2}$. In the following we assume for simplicity no 
Charge-Parity (CP) violation in the scalar sector. Then, apart from the recently observed (CP-even) Higgs boson $h$, 
the 2HDM physical spectrum contains a charged scalar $H^{\pm}$, a CP-even scalar $H_0$, and a CP-odd scalar $A_0$.
After fixing the electroweak (EW) vacuum expectation value (vev) $v = 246$ GeV and the Higgs mass $m_h = 125$~GeV, the
remaining parameters in the scalar potential are: the masses $m_{H_0}$, $m_{A_0}$, $m_{H^{\pm}}$, two angles $\beta$ and $\alpha$
and a dimensionful parameter $\mu$. Here the angle $\alpha$ is defined such that when $\alpha = \beta$, the state $H_0$ decouples from gauge bosons, and $h$ 
has SM-like properties, known as the \emph{alignment limit} (see \cite{Branco:2011iw,Gunion:2002zf} for a review of 2HDMs, 
and \cite{Dorsch:2013wja} for details on the definition of parameters and the conventions chosen).

In order to study the strength of the EWPT in 2HDMs, a scan over the new physics parameters $m_{H_0}$, $m_{A_0}$, $m_{H^{\pm}}$, $\tan\beta$, 
$\alpha - \beta$ and $\mu$ is performed. The Yukawa type of 2HDM considered is irrelevant for the EWPT (all Types couple in the same way to the 
top quark), while experimental constraints do differ between Types. We choose a Type I 2HDM, which is less constrained than a Type II, and thus provides a better gauging of the impact 
of a first order EWPT on the 2HDM parameter space. The scan is interfaced to 2HDMC~\cite{Eriksson:2009ws} and HiggsBounds~\cite{Bechtle:2013wla} to select points in parameter 
space that satisfy stability, unitarity, perturbativity, EW precision constraints and collider bounds. 
Flavour constraints (coming mainly from $b\rightarrow s\gamma$ for Type I 2HDM \cite{Mahmoudi:2009zx}) and constraints 
from measured Higgs signal strengths on $\tan\beta$ and $\alpha-\beta$ (see e.g. \cite{Celis:2013rcs}) are also included.

\begin{figure}[ht]
\begin{center}
	\includegraphics[width=0.63\textwidth, clip]{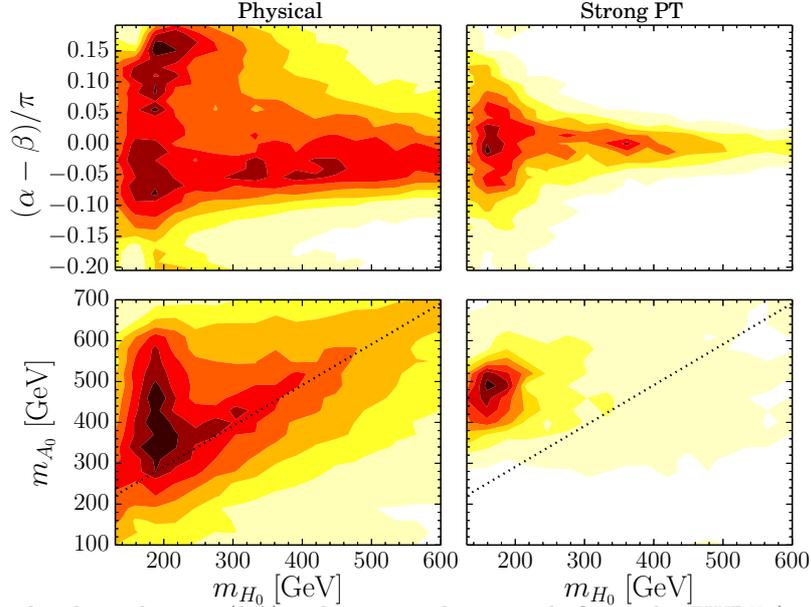}
	\vspace{-7mm}
	\caption{\small  Heat-maps for the physical region (\emph{left}) and region with a strongly first 
	order EWPT (\emph{right}). \emph{Top}: ($m_{H_0}, \alpha - \beta$)-plane. \emph{Bottom}: ($m_{H_0}, m_{A_0}$)-plane. The 
	dotted-black line corresponds to $m_{A_0} = m_{H_0} + m_{Z}$.}
	\vspace{-7mm}
	\label{fig:EWPT}
	\end{center}
\end{figure}

Point in our scan satisfying the above constraints are called \emph{physical points}. For them, the strength of the EWPT is computed via the thermal 1-loop effective 
potential (see \cite{Dorsch:2013wja} for details). In Figure~\ref{fig:EWPT} we show heat-maps of physical points (\emph{left}) and points with a strongly first order 
EWPT (\emph{right}) in the planes ($m_{H_0}, \alpha - \beta$) and ($m_{H_0}, m_{A_0}$). Altogether, a strong EWPT as needed for Electroweak Baryogenesis strongly 
favours a rather heavy CP-odd scalar $A_0$ ($m_{A_0} > 300$ GeV), together with a large mass splitting $m_{A_0} - m_{H_0} \gtrsim v$, as shown in 
Figure~\ref{fig:EWPT} (bottom). It also favours the light Higgs $h$ to have SM-like properties, {\it i.e.} small $\alpha-\beta$ and 
moderate $\tan\beta$ \cite{Dorsch:2014qja,Dorsch:2013wja}, the range of $\alpha-\beta$ leading to a strong EWPT shrinking as the 
CP-even state $H_0$ becomes heavier. 

\section{The Decay $A_0 \rightarrow Z\,H_0$}
\label{section3}
\vspace{-3mm}

\begin{figure}[ht]
\begin{center}
	\includegraphics[width=0.45\textwidth, clip]{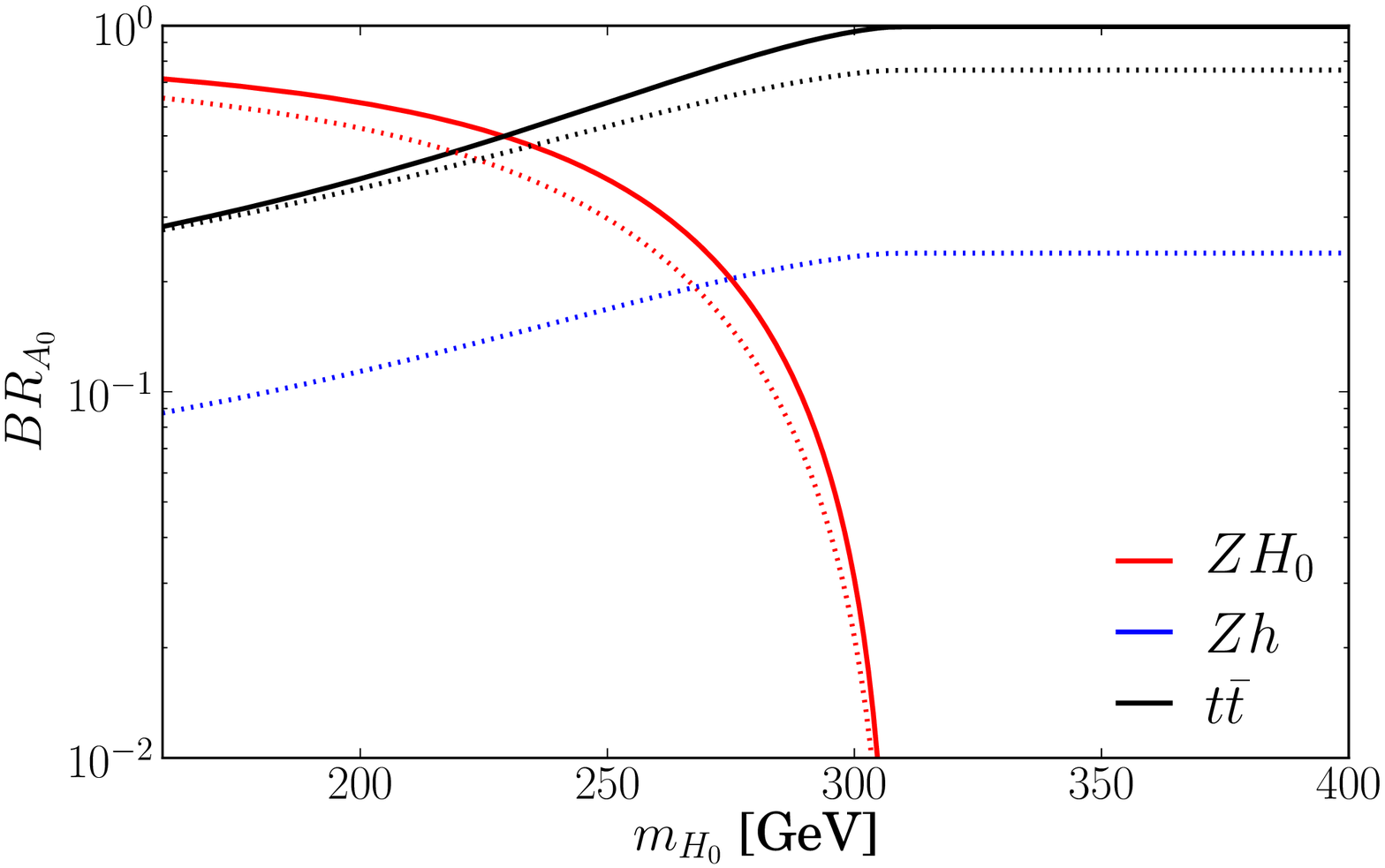}\hspace{4mm}
	\includegraphics[width=0.45\textwidth, clip]{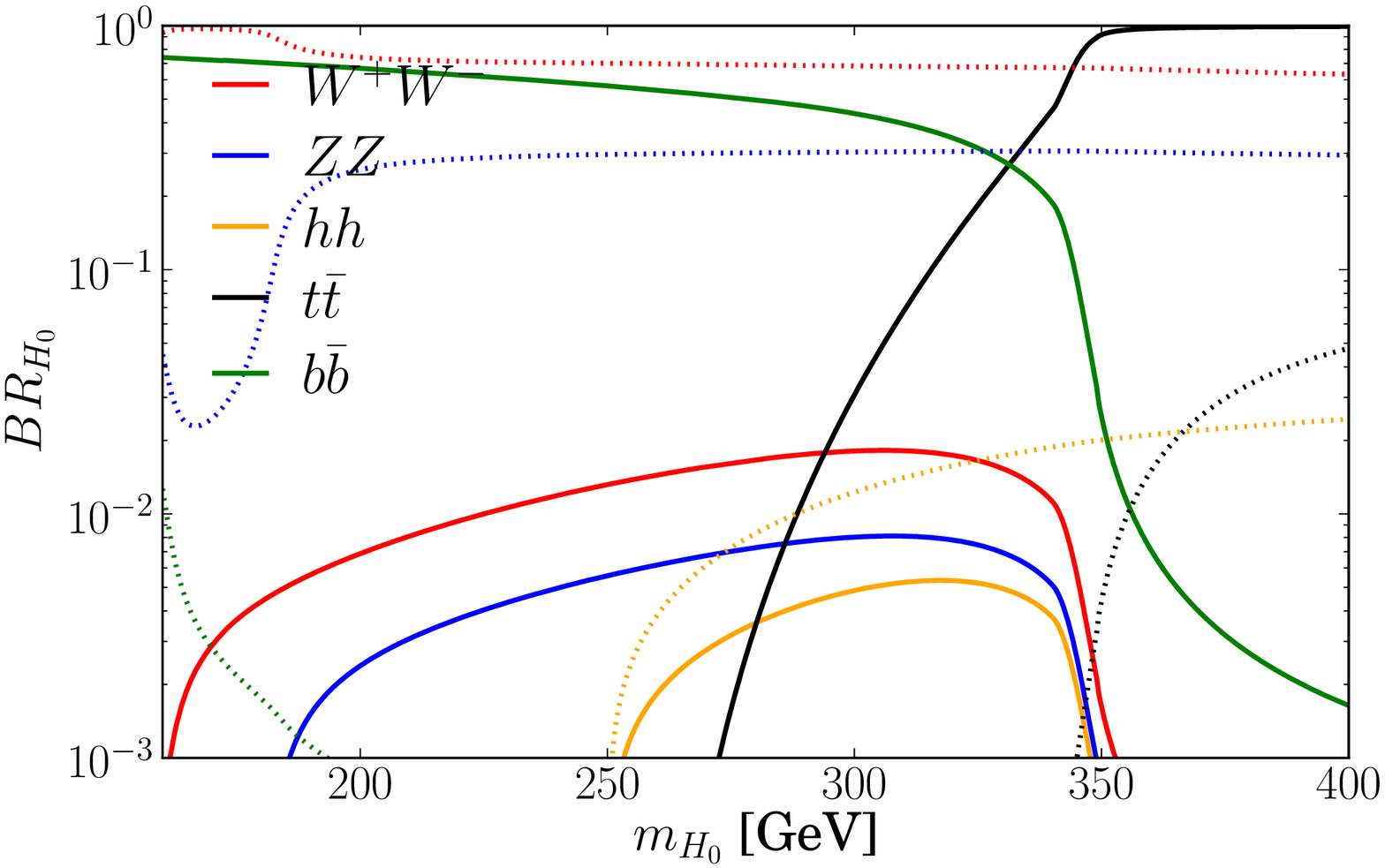}
    \vspace{-2mm}
	\caption{\small \emph{Left}: Main Branching Ratios of the CP-odd scalar $A_0$ as a function of $m_{H_0}$ for $m_{A_0} = m_{H^{\pm}} = 400$~ GeV, 
	$\tan\beta = 2$, $\mu = 100$~GeV, $\alpha - \beta = 0.001\pi$ (benchmark A, solid lines) and $\alpha - \beta = 0.1\pi$ (benchmark B, dotted lines). \emph{Right}:  
	Main Branching Ratios of $H_0$ as a function of $m_{H_0}$ (same benchmark parameters as in \emph{Left}).}
	\vspace{-5mm}
	\label{fig:Br}
	\end{center}
\end{figure}

The results of the previous section point towards the $A_0 \rightarrow Z H_0$ decay channel (see also \cite{Coleppa:2014hxa}) as a characteristic signature of 2HDMs with a strong EWPT, to be 
searched for at the upcoming 14 TeV run of the LHC. This decay is strongly favoured both by the large phase space available and by the 
$c_{\alpha-\beta}$ dependence of the coupling $g_{A_0ZH_0}$, unsuppressed in the alignment limit. This is in contrast with the decay 
$A_0 \rightarrow Z h$, which vanishes in that limit (since $g_{A_0Zh} \sim s_{\alpha-\beta}$), and is therefore suppressed in this scenario (see Figure \ref{fig:Br}). 
The competing decay channels would then be $A_0 \rightarrow t\bar{t}$ and possibly $A_0 \rightarrow W^{\pm} H^{\mp}$. The former is subdominant 
for $m_{A_0}-m_{H_0} \gtrsim v$ (Figure \ref{fig:Br}). The presence of the latter depends on the splitting $m_{A_0}-m_{H^{\pm}}$. 
EW precision observables require $H^{\pm}$ to be close in mass to either $H_0$ or $A_0$ \cite{Grimus:2007if},
which makes $A_0 \rightarrow W^{\pm} H^{\mp}$ either kinematically forbidden or similar to $A_0 \rightarrow Z H_0$,
and here we assume for simplicity $m_{H^{\pm}} \sim m_{A_0}$. 

In the following we analyze two prototypical scenarios which feature $\mu = 100$~GeV, $\mathrm{tan}\beta = 2$ and $m_{A_0} = m_{H^{\pm}} = 400$~GeV, and with 
$(\alpha - \beta) = 0.001\,\pi$ (Benchmark A) and $(\alpha - \beta) = 0.1\,\pi$ (Benchmark B), as shown in Figure \ref{fig:Br}.
These two benchmarks characterize the two dominant alternatives for the subsequent decay of $H_0$: $H_0 \rightarrow b \bar{b}$ dominates
very close to the alignment limit, while away from it $H_0\rightarrow W^+W^-$ (and to a lesser extent $H_0\rightarrow Z Z$) is the dominant
decay mode.

This discussion highlights the fact that for 2HDMs with a strongly first order EWPT, the corresponding ``smoking gun" signature at LHC will 
either be $p p \rightarrow A_0 \rightarrow Z H_0 \rightarrow \ell \ell b \bar{b}$ or $p p \rightarrow A_0 \rightarrow Z H_0 \rightarrow \ell \ell W^+W^-$, depending 
on how close the 2HDM is to the alignment limit. 

\section{LHC Search for $A_0$ in $\ell\ell\,b\bar{b}$ and $\ell\ell\,W^{+}W^{-}$}

We now analyze the search prospects in the $\ell \ell b \bar{b}$ and $\ell \ell W^+W^-$ channels at the 14 TeV run of the LHC using the defined benchmarks A and B from 
section \ref{section3} and considering in both cases a mass for the CP even scalar $m_{H_0} = 180$~GeV. 

\begin{figure}[ht]
\begin{center}
	\includegraphics[width=0.48\textwidth, clip]{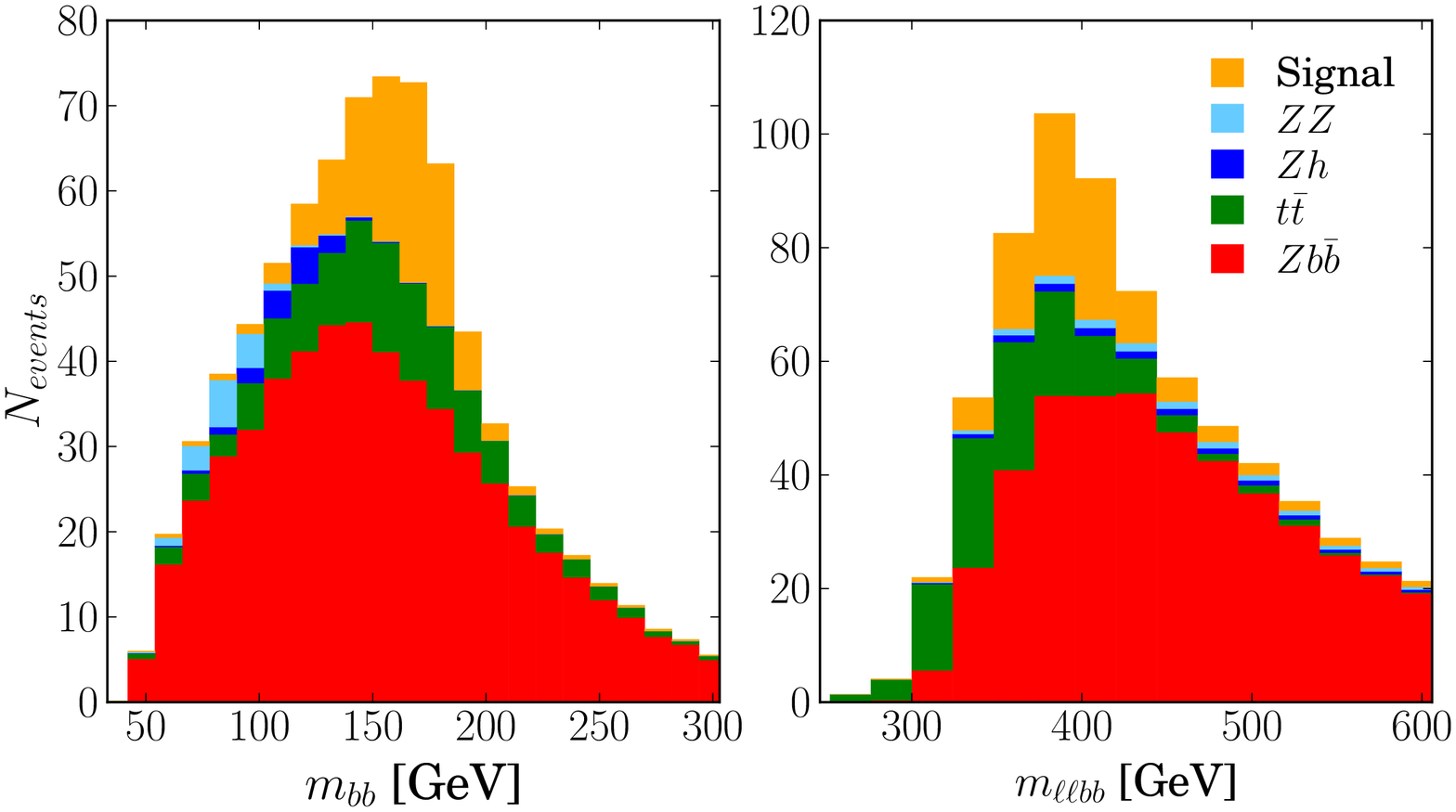}\hspace{3mm}
	\includegraphics[width=0.48\textwidth, clip]{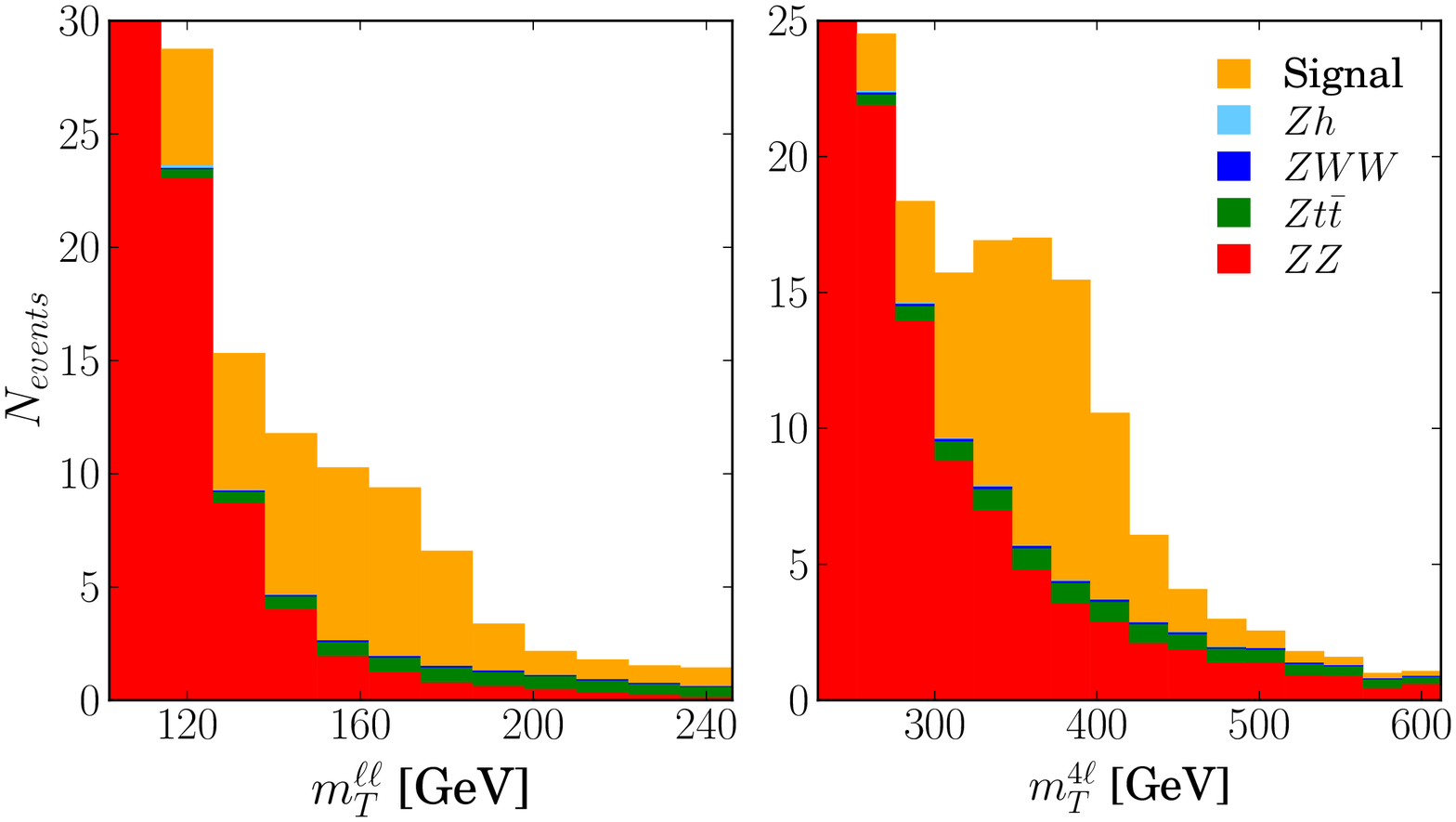}
	\vspace{-5mm}
	\caption{\small Left: $m_{bb}$ (\emph{left}) and $m_{\ell\ell bb}$ (\emph{right}) distributions after analysis cuts, 
	with the various contributions stacked (for an integrated luminosity of $\mathcal{L} = 20\,\, \mathrm{fb}^{-1}$). Right: 
	$m^{\ell\ell}_{T}$ (\emph{left}) and $m^{4\ell}_{T}$ (\emph{right}) distributions after event selection, with the 
	various contributions stacked (for an integrated luminosity of $\mathcal{L} = 60\, \,\mathrm{fb}^{-1}$).}
	\vspace{-5mm}
	\label{fig:dist1}
	\end{center}
\end{figure}

Concentrating first on Benchmark A, which corresponds to the $\ell \ell b \bar{b}$ final state, the two main SM backgrounds are: (i) $Z b\bar{b}$ production 
(with $Z \rightarrow \ell \ell$), (ii) QCD $t\bar{t}$ production (with $t\bar{t} \rightarrow b W^{+} \bar{b} W^{-} \rightarrow b \ell^{+} \nu_{\ell} \bar{b} \ell^{-} \bar{\nu}_{\ell}$), 
while the rest of potential backgrounds (e.g. $Z Z$ production and associated production of a Higgs $h$ and a $Z$ boson) are found to be practically 
negligible \cite{Dorsch:2014qja}. Regarding our analysis, we implement the Type-I 2HDM in {\sc FeynRules} \cite{Christensen:2008py} 
and use {\sc MadGraph5$\_$aMC$@$NLO} \cite{Alwall:2011uj} to generate both signal and background analysis samples, which are then passed on to {\sc Pythia} 
\cite{Sjostrand:2007gs} and {\sc Delphes} \cite{deFavereau:2013fsa} for parton showering, hadronization and a detector simulation.
For event selection we require the presence of two isolated same flavour (SF) leptons in the final state with $P^{\ell_1}_{T} > 40$, $P^{\ell_2}_{T} > 20$ and $\left| \eta_{\ell} \right| < $ 2.5 
(2.7) for electrons (muons), together with two b-tagged jets in the event with $P^{b_1}_{T} > 40$, $P^{b_2}_{T} > 20$ and $\left| \eta_{b} \right| < $ 2.5
(see \cite{Dorsch:2014qja} for details), and the subsequent cut-flow analysis is presented in Table \ref{Table1}. 
We define the signal region as $m_{bb} = (m_{H_0} - 20) \pm 30$ GeV and $m_{\ell\ell bb} = (m_{A_0} - 20) \pm 40$ GeV 
(small b-jet energy loss due to showering is expected), and show the $m_{bb}$ and $m_{\ell\ell bb}$ distributions after cuts in Figure \ref{fig:dist1} (Left)
(various contributions stacked and for an integrated luminosity of $\mathcal{L} = 20\, \mathrm{fb}^{-1}$). The results 
from Table \ref{Table1} show that a discovery value $S/\sqrt{S+B} = 5$ may be obtained already with $\mathcal{L}\sim 15-20$ fb$^{-1}$ in the limit that 
only statistical uncertainties are important.

\begin{table}[ht]
\caption{\small Event selection (see section III) and background reduction in the $\ell \ell b \bar{b}$ final state. Cross section (in fb) is shown after 
successive cuts for the signal $A_0 \rightarrow Z H_0 \rightarrow \ell \ell b \bar{b}$ and the dominant backgrounds $t \bar{t}$ and $Z b \bar{b}$.}
\label{Table1}
\vspace{2mm}
\begin{tabular}{c| c | c | c | }
& Signal & $t\bar t$ & $Z\, b\bar b$ \\
\hline 
&  &  &  \\ [-2ex]
Event selection& 14.6 & 1578 & 424  \\ [0.5ex]
$80$ $< m_{\ell\ell} < 100$ GeV& 13.1 & 240 & 388  \\ [0.5ex]
$\begin{array}{c}
H_T^\mathrm{bb} > 150 \,\mathrm{GeV} \\
H_T^\mathrm{\ell\ell bb} > 280\, \mathrm{GeV}
 \end{array}$
& 8.2 & 57 & 83  \\ [2ex]
$\Delta R_{bb} < 2.5$, $\Delta R_{\ell\ell} < 1.6$& 5.3 & 5.4 & 28.3  \\ [0.5ex]
\hline &  &  &  \\ [-2ex]
$m_{bb}$, $m_{\ell\ell bb}$ signal region& 3.2 & 1.37 & 3.2  \\ [0.5ex]
\hline
\end{tabular}
\end{table}

Allowing a departure from the alignment limit (our Benchmark B) the decay channel 
$H_0 \rightarrow W^+W^- \to \ell \nu_{\ell} \ell \nu_{\ell}$, $Z \to \ell' \ell'$  
provides the best discovery prospects ($H_0 \rightarrow Z Z$, leading to a final state $Z Z Z \to \ell \ell \ell' \ell' j j$ has been 
considered in \cite{Coleppa:2014hxa}). The main background is diboson ($Z Z$) production with $Z Z \rightarrow \ell \ell \ell' \ell'$
(other backgrounds like $Z t \bar{t}$, $Z W W$ and $Z h$ are very small even before cuts). For event selection, we require the 
presence of four isolated leptons in the final state with $P^{\ell_1}_{T} > 40$ GeV, $P^{\ell_2,\ell_3,\ell_4}_{T} > 20$ GeV, and require that 
one lepton pair (opposite sign, same flavour (SF)) reconstructs $m_Z$ within $20$ GeV. 
The Leading Order cross sections at LHC 14 TeV after event selection for the signal and $Z Z$ background are respectively $0.93$ fb and $5.6$ fb.
We note that a $Z$-veto on the remaining lepton pair would greatly suppress the $Z Z$ background, as would a veto on SF leptons (this at 
the expense of reducing the signal by a factor 2). Defining the transverse mass variables $m^{\ell\ell}_{T}$ and $m^{4\ell}_{T}$
\begin{equation}
\left(m^{\ell\ell}_{T}\right)^2 =  \left(\sqrt{p^2_{T,\ell\ell}+m^2_{\ell\ell}}  + \slash\hspace{-2mm} p_{T}\right)^2 - \left(\vec{p}_{T,\ell\ell} + 
\slash\hspace{-2mm}\vec{p}_{T}\right)^2 \quad \quad 
m^{4\ell}_{T} = \sqrt{p^2_{T,\ell'\ell'}+m^2_{\ell'\ell'}} + \sqrt{p^2_{T,\ell\ell}+\left(m^{\ell\ell}_{T}\right)^2}
\end{equation}

($\ell'\ell'$ are the two SF leptons most closely reconstructing $m_Z$), a signal region of $m^{4\ell}_{T} > 260$ GeV 
(see Figure \ref{fig:dist1}, Right) allows to extract a clean signal \cite{Dorsch:2014qja}. Our final signal cross section is 1.41 fb, which compared 
to a background of 1.7 fb reaches a significance of 5 with $\mathcal{L}\sim 60$ fb$^{-1}$.

\vspace{2mm}

Altogether, this analysis highlights that the decay $A_0 \rightarrow Z H_0$, being a `smoking gun' signature of 2HDM scenarios with a strongly first order EWPT,
can be probed at the 14 TeV run of LHC in either of the two dominant final states, thus providing a powerful probe of the EWPT and EW Cosmology at the LHC.

\begin{acknowledgments}
I am deeply indebted to Glauber Dorsch, Stephan Huber and Ken Mimasu, from which I learned a lot by working together, 
and who made our collaboration really fun. I also want to thank Veronica Sanz and Michael Ramsey-Musolf for useful discussions and comments,
and the organizers of HPNP2015 for their great hospitality during the conference.
The work of J.M.N. is supported by the People Programme (Marie curie Actions) of the European Union Seventh Framework Programme (FP7/2007-2013) 
under REA grant agreement PIEF-GA-2013-625809. 
\end{acknowledgments}

\bigskip 

\begin{thebibliography}{99} 


\bibitem{EWBG} 
  N.~Turok and J.~Zadrozny,
  Nucl.\ Phys.\ B {\bf 358}, 471 (1991);   
  J.~M.~Cline, K.~Kainulainen and A.~P.~Vischer,
  Phys.\ Rev.\ D {\bf 54}, 2451 (1996)
  [hep-ph/9506284];
  L.~Fromme, S.~J.~Huber and M.~Seniuch,
  JHEP {\bf 0611} (2006) 038
  [hep-ph/0605242];
  J.~M.~Cline, K.~Kainulainen and M.~Trott,
  JHEP {\bf 1111}, 089 (2011)
  [arXiv:1107.3559 [hep-ph]].    

\bibitem{Kajantie:1995kf}
  K.~Kajantie, M.~Laine, K.~Rummukainen and M.~E.~Shaposhnikov,
  Nucl.\ Phys.\ B {\bf 466} (1996) 189
  [hep-lat/9510020]; 
  K.~Kajantie, M.~Laine, K.~Rummukainen and M.~E.~Shaposhnikov,
  Phys.\ Rev.\ Lett.\  {\bf 77} (1996) 2887
  [hep-ph/9605288].  
  
\bibitem{Dorsch:2014qja}
  G.~C.~Dorsch, S.~J.~Huber, K.~Mimasu and J.~M.~No,
  Phys.\ Rev.\ Lett.\  {\bf 113} (2014) 21,  211802
  [arXiv:1405.5537 [hep-ph]].  
  
\bibitem{Branco:2011iw} 
  G.~C.~Branco, P.~M.~Ferreira, L.~Lavoura, M.~N.~Rebelo, M.~Sher and J.~P.~Silva,
  Phys.\ Rept.\  {\bf 516}, 1 (2012)
  [arXiv:1106.0034 [hep-ph]].  

\bibitem{Gunion:2002zf}
  J.~F.~Gunion and H.~E.~Haber,
  Phys.\ Rev.\ D {\bf 67} (2003) 075019
  [hep-ph/0207010].  
  
\bibitem{Dorsch:2013wja}
  G.~C.~Dorsch, S.~J.~Huber and J.~M.~No,
  JHEP {\bf 1310} (2013) 029
  [arXiv:1305.6610 [hep-ph]].  
 
 
\bibitem{Eriksson:2009ws} 
  D.~Eriksson, J.~Rathsman and O.~Stal,
  Comput.\ Phys.\ Commun.\  {\bf 181}, 189 (2010)
  [arXiv:0902.0851 [hep-ph]].  
  
\bibitem{Bechtle:2013wla}
  P.~Bechtle, O.~Brein, S.~Heinemeyer, O.~Stål, T.~Stefaniak, G.~Weiglein and K.~E.~Williams,
  Eur.\ Phys.\ J.\ C {\bf 74} (2014) 2693
  [arXiv:1311.0055 [hep-ph]]. 
 
  
\bibitem{Mahmoudi:2009zx}
  F.~Mahmoudi and O.~Stal,
  Phys.\ Rev.\ D {\bf 81} (2010) 035016
  [arXiv:0907.1791 [hep-ph]];
  T.~Hermann, M.~Misiak and M.~Steinhauser,
  JHEP {\bf 1211} (2012) 036
  [arXiv:1208.2788 [hep-ph]].
  
  
\bibitem{Celis:2013rcs} 
  A.~Celis, V.~Ilisie and A.~Pich,
  JHEP {\bf 1307}, 053 (2013)
  [arXiv:1302.4022 [hep-ph]];
  M.~Krawczyk, D.~Sokolowska and B.~Swiezewska,
  J.\ Phys.\ Conf.\ Ser.\  {\bf 447}, 012050 (2013)
  [arXiv:1303.7102 [hep-ph]];
  B.~Grinstein and P.~Uttayarat,
  JHEP {\bf 1306}, 094 (2013)
  [Erratum-ibid.\  {\bf 1309}, 110 (2013)]
  [arXiv:1304.0028 [hep-ph]];
  C.~-Y.~Chen, S.~Dawson and M.~Sher,
  Phys.\ Rev.\ D {\bf 88}, 015018 (2013)
  [arXiv:1305.1624 [hep-ph]]; 
  O.~Eberhardt, U.~Nierste and M.~Wiebusch,
  JHEP {\bf 1307}, 118 (2013)
  [arXiv:1305.1649 [hep-ph]]; 
  G.~Belanger, B.~Dumont, U.~Ellwanger, J.~F.~Gunion and S.~Kraml,
  Phys.\ Rev.\ D {\bf 88} (2013) 075008
  [arXiv:1306.2941 [hep-ph]];
  B.~Dumont, J.~F.~Gunion, Y.~Jiang and S.~Kraml,
  Phys.\ Rev.\ D {\bf 90} (2014) 035021
  [arXiv:1405.3584 [hep-ph]];
  M.~Gorbahn, J.~M.~No and V.~Sanz,
  arXiv:1502.07352 [hep-ph].
 
 
 \bibitem{Grimus:2007if} 
  W.~Grimus, L.~Lavoura, O.~M.~Ogreid and P.~Osland,
  J.\ Phys.\ G {\bf 35}, 075001 (2008)
  [arXiv:0711.4022 [hep-ph]]; 
  Nucl.\ Phys.\ B {\bf 801}, 81 (2008)
  [arXiv:0802.4353 [hep-ph]]. 
 
\bibitem{Coleppa:2014hxa}
  B.~Coleppa, F.~Kling and S.~Su,
  JHEP {\bf 1409} (2014) 161
  [arXiv:1404.1922 [hep-ph]]. 
  

\bibitem{Christensen:2008py} 
  N.~D.~Christensen and C.~Duhr,
  Comput.\ Phys.\ Commun.\  {\bf 180}, 1614 (2009)
  [arXiv:0806.4194 [hep-ph]];
  C.~Degrande, C.~Duhr, B.~Fuks, D.~Grellscheid, O.~Mattelaer and T.~Reiter,
  Comput.\ Phys.\ Commun.\  {\bf 183}, 1201 (2012)
  [arXiv:1108.2040 [hep-ph]].  
  
\bibitem{Alwall:2011uj} 
  J.~Alwall, M.~Herquet, F.~Maltoni, O.~Mattelaer and T.~Stelzer,
  JHEP {\bf 1106}, 128 (2011)
  [arXiv:1106.0522 [hep-ph]];  
J.~Alwall, R.~Frederix, S.~Frixione, V.~Hirschi, F.~Maltoni, O.~Mattelaer, H.-S.~Shao and T.~Stelzer {\it et al.},
  JHEP {\bf 1407} (2014) 079
  [arXiv:1405.0301 [hep-ph]].  
  
\bibitem{Sjostrand:2007gs} 
  T.~Sjostrand, S.~Mrenna and P.~Z.~Skands,
  Comput.\ Phys.\ Commun.\  {\bf 178}, 852 (2008)
  [arXiv:0710.3820 [hep-ph]].
  
\bibitem{deFavereau:2013fsa} 
  J.~de Favereau {\it et al.}  [DELPHES 3 Collaboration],
  JHEP {\bf 1402}, 057 (2014)
  [arXiv:1307.6346 [hep-ex]].  

 

\end{thebibliography}

\end{document}